\documentclass[10pt,letterpaper]{article}
\usepackage[top=0.85in,left=2.75in,footskip=0.75in]{geometry}

\usepackage{amsmath,amssymb}

\usepackage{changepage}

\usepackage[utf8x]{inputenc}

\usepackage{textcomp,marvosym}

\usepackage{cite}

\usepackage{nameref,hyperref}

\usepackage[right]{lineno}

\usepackage{microtype}
\DisableLigatures[f]{encoding = *, family = * }

\usepackage[table]{xcolor}

\usepackage{array}

\newcolumntype{+}{!{\vrule width 2pt}}

\newlength\savedwidth

\raggedright
\usepackage[aboveskip=1pt,labelfont=bf,labelsep=period,justification=raggedright,singlelinecheck=off]{caption}

\makeatletter
\renewcommand{\@biblabel}[1]{\quad#1.}
\makeatother

\date{}

\usepackage{lastpage,fancyhdr,graphicx}
\usepackage{epstopdf}
\fancyheadoffset[L]{2.25in}
\fancyfootoffset[L]{2.25in}

\begin{document}
\vspace*{0.2in}

\begin{flushleft}
{\Large
\textbf\newline{Cooperative dynamics of Neighborhood Economic Status in cities} 
}
\newline
\\
Anand Sahasranaman\textsuperscript{1\textcurrency*},
Henrik Jeldtoft Jensen\textsuperscript{1\textcurrency}
\\
\bigskip
\textbf{1} Department of Mathematics and Centre for Complexity Science, Imperial College London, London, United Kingdom
\\
\bigskip

%
%
\textcurrency Current Address: Centre for Complexity Science, 12th Floor EEE Building, Imperial College London, London, SW7 2AZ, United Kingdom 


\bigskip
* a.sahasranaman15@imperial.ac.uk

\end{flushleft}
\section*{Abstract}
We significantly extend our earlier variant of the Schelling model, incorporating a neighborhood Potential function as well as an agent wealth gain function to study the long term evolution of the economic status of neighborhoods in cities. We find that the long term patterns of neighborhood relative economic status (RES) simulated by this model reasonably replicate the empirically observed patterns from American cities. Specifically, we find that larger fractions of rich and poor neighborhoods tend to, on average, retain status for longer than lower- and upper-middle wealth neighborhoods. The use of a Potential function that measures the relative wealth of neighborhoods as the basis for agent wealth gain and agent movement appears critical to explaining these emergent patterns of neighborhood RES. This also suggests that the empirically observed RES patterns could indeed be universal and that we would expect to see these patterns repeated for cities around the world. Observing RES behavior over even longer periods of time, the model predicts that the fraction of poor neighborhoods retaining status remains almost constant over extended periods of time, while the fraction of middle-wealth and rich neighborhoods retaining status reduces significantly over time, tending to zero.

\section*{Introduction}
Cities are increasingly being seen as systems that are potentially amenable to exploration through the tools of complex systems science and statistical physics. This includes work on processes underlying the the emergence of inequitous outcomes such as segregation.

The Schelling model~\cite{bib1} explained the emergence of racial segregation in terms of interactions of individual preferences, with even small preferences for like neighbors at an individual level leading to emergent segregated patterns at a collective level. The Schelling model has been found robust across parameter specifications relating to neighborhood shape, size, heterogeneity of agent preferences, and agent choice functions~\cite{bib2a}~\cite{bib2b}~\cite{bib2c}~\cite{bib2d}~\cite{bib2e}~\cite{bib2f}, and has been widely applied to agent-based simulations in sociology~\cite{bib2g}~\cite{bib2h}~\cite{bib2i} and economics~\cite{bib2j}. The tools of statistical mechanics have been used to study the underlying phase transition occurring in the model, exploring the interface between segregation and de-segregation~\cite{bib2k}~\cite{bib2l}~\cite{bib2m}~\cite{bib2n}. The Schelling model has also been extended to agents with a wealth or income attribute sampled from a continuous distribution~\cite{bib2g}~\cite{bib2i}. Benenson, Hatna, and Or~\cite{bib2g} use the Schelling model to test the emergence of income-based segregation, and also suggest that the introduction of a small number of highly tolerant agents could significantly increase the heterogeneity of residential patterns. Benard and Willer~\cite{bib2i}  find significant wealth based segregation with increasing correlation between status and wealth, and with increasing endogeneity in housing prices.

Previously, we had modeled the phenomenon of wealth based segregation~\cite{bib3aa} using a modification of the classic Schelling model~\cite{bib1} and found a sharp transition from segregated to mixed equilibrium when only a small fraction of agents were allowed to move into neighborhoods they could not afford, thus agreeing with the empirical finding of Benenson, Hatna, and Or~\cite{bib2g}, while offering an alternate interpretation based not on high tolerance levels of a few agents but on allowing a small proportion of agents to contravene the wealth thresholds imposed by neighborhoods. While this model provides a mechanism to explore segregation and processes that enable its reversal, it does not provide insight into the long-term evolution of economic status of neighborhoods. Therefore, we now seek to extend the model with the objective of modeling the evolution of economic status of neighborhoods over long sweeps of time.   

The processes of segregation and neighborhood decline are of significant public policy importance because of their detrimental effects on long-term socio-economic outcomes, as evidenced by a large body of literature in sociology and economics ~\cite{bib5}~\cite{bib6}~\cite{bib7}. Economic homogeneity is found to have adverse impacts on education attainment~\cite{bib5}, civic participation~\cite{bib6}, and the risk of mortality~\cite{bib7} of poor citizens.  

Rosenthal and Ross~\cite{bib10} review the available evidence on change and persistence of economic status of neighborhoods in cities over long time periods. The central finding is that though many locations exhibit extreme persistence in economic status, it is also common for locations to experience change in economic status. The authors reiterate that long sweeps of time are necessary to appreciate the fact that changes in economic status of neighborhoods are common. Rosenthal~\cite{bib11} shows that over half of the neighborhoods in core areas of 35  US cities were of markedly different economic status in 2000 compared to 1950. He finds that change in economic status is common and that two-thirds of the neighborhoods in studied American cities changed economic status in this period. However, certain types of neighborhoods were found to have changed economic character with greater likelihood than others - the economic status of 34.2\% of neighborhoods in the first quartile (of economic status) and 43.9\% of neighborhoods in the fourth quartile over 1950-2000 did not change, while only 26.4\% of the second quartile and 26.9\% of the third quartile neighborhoods retained their economic status in the same time period. This variation in economic status is explained through a few significant mechanisms such as aging housing stocks and redevelopments, as well as local externalities associated with social interactions.

Breuckner and Rosenthal~\cite{bib11a} argue that as housing which is initially built for richer families decays over time and the rich move out to newer housing that is built in an outwardly growing city, the older decaying housing is passed on to lower income populations. Consequently, as the decaying housing is redeveloped, the rich move back in to the city resulting in cycles of change in economic status of neighborhoods. 

Schelling's model~\cite{bib1} provides a dynamic explanation of economic segregation. Economic segregation is reinforced by preferential attachment, which refers to processes where some attribute (such as wealth in our case) is distributed among agents and grows in accordance with how much they already have, yielding a 'rich-get-richer' dynamic~\cite{bib11b}. Bernabeu~\cite{bib12} finds that in a system with two neighborhoods where the cost of skill acquisition decreases with increasing skills because of positive peer effects (or preferential attachment), if communities were to form on the basis of investment in skills, then this leads to higher levels of skill inequality and reinforces stratification of high and low incomes in different neighborhoods. There is also strong evidence of 'tipping' behavior in American cities where, as the minority household share increases to 5-20\%, majority households flee and the neighborhood tips to become a minority neighborhood~\cite{bib12a}~\cite{bib12b}. Ferrerya~\cite{bib13} finds that school vouchers reduce income segregation by weakening the link between the household's residential location and the quality of schools for children. Therefore, socio-demographic variables such as education and skills appear to have strong explanatory powers for the rate at which neighborhoods' economic status transition up and down. Rosenthal~\cite{bib11} finds that neighborhoods that have positive socio-demographic attributes tend to have high economic status even as housing stock ages. It also emerges that higher home ownership rates reduce the likelihood that the economic status of the neighborhood goes down, and because home ownership is more prevalent in high income locations, such neighborhoods tend to retain their economic status over time.

Consequently, the evidence on the impact of neighborhood effects over extended time periods reveals that change in economic status of neighborhoods is very common and that while indeed all types of neighborhoods change status, rich and poor neighborhoods tend to retain character more often than middle wealth neighborhoods. This change in economic status of neighborhoods is also found to be driven by the emergence and persistence of wealth based segregation.

Given these empirical observations, our objective is to extend the model we presented in Sahasranaman and Jensen~\cite{bib3aa}, by incorporating temporal wealth change dynamics and thereby study patterns in changes of neighborhood economic status over long time sweeps and their relation to the extent of wealth segregation. We explore whether the empirically observed patterns are replicable in the model and if indeed there is a case to be made for the universality of these patterns. 

\section*{Model Definition and Specifications}
We seek to understand the evolution of neighborhood economic status in a city composed of $M$ neighborhoods. Each of the neighborhoods, $i$ ($i$ $\in$ $1$,...,$M$), is populated by $P(i)$ agents (households) and each agent has a single attribute - wealth - which is the driver of the model dynamics. Initially, agents are distributed equally across each of the M neighborhoods, such that $P(i) = P(j)$ for $1 \le i,j \le M$. Empirical work has shown that real world wealth distributions tend to be heavy-tailed~\cite{bib14}~\cite{bib15}, and could potentially represent log-normal or stretched exponential functions. Therefore, for the purpose of this model we sample agent wealths from a log-normal (LN) distribution. In order to simulate different initial wealth configurations, we draw wealths from two log-normal distributions - LN ($\mu$ = 0, $\sigma$ = 0.25) and LN ($\mu$ = 0, $\sigma$ = 0.5). Additionally, given the fact that cities around the world are economically segregated to different extents~\cite{bib15a}~\cite{bib15b}, we create initial wealth configurations representing different levels of segregation. We do this by ordering the realized log-normal wealth distribution in ascending order, and creating $M$ $P$-sized blocks of sorted agent wealths. Each of these contiguous blocks is assigned to a neighborhood, until each of the $M$ neighborhoods is populated by one block of $P$ agents. This leads to configurations where, on average, 80\% of city wealth is owned by between 72 \% and 73\% of the agents for the LN ($\mu$ = 0, $\sigma$ = 0.25) distribution and by between 53 \% and 60\% of the agents for the LN ($\mu$ = 0, $\sigma$ = 0.5) distribution. The total wealth of each neighborhood $i$ is the sum of individual agent wealths and is denoted by $W\textsubscript{tot}(i)$. 

The dynamics are driven by decisions of the individual agents (households) in the city. At each iteration, a random household is sampled, and this household makes a probabilistic choice of either staying in its current neighborhood or attempting to move to an alternate neighborhood. This choice is based on an evaluation of the wealth of the agent's current neighborhood in relation to other neighborhoods in the city. The use of such a comparative metric is supported by the evidence on impact of neighborhoods on long term social and economic outcomes of households. In addition to the research cited previously, recent findings show that neighborhoods affect inter-generational mobility and that outcomes of children who move into a better neighborhood improve linearly with the time spent in that area~\cite{bib16}. There is also a significant positive impact on employment and income levels of children moving out of disadvantaged neighborhoods~\cite{bib17}. We propose a metric - Potential ($U$) - to assess the satisfaction of an agent with its present neighborhood. We define the Potential ($U\textsubscript{C}$) of a specific neighborhood ($C$) as a function that computes the economic status of $C$ relative to the economic status of all the $M$ neighborhoods in the city. Specifically, we define $U\textsubscript{C}$ as the ratio of the total wealth of neighborhood $C$ relative to the average total neighborhood wealth of the city (Eq~\ref{eq:pot}): 

\begin{eqnarray}
\label{eq:pot}
	\mathrm{U_C} \equiv \frac{W_{tot}(C)}{\sum_{j=1}^{M}W_{tot}(j)/M}
\end{eqnarray}

$U < 1$ implies that the agent is in a neighborhood with total wealth lower than the the total wealth of an average neighborhood in the city. $U > 1$ implies that the agent is in a neighborhood with higher wealth than the average city neighborhood. $U = 1$ implies that the neighborhood is on par with the average city neighborhood. Given $U$, the probability of a household staying in its current neighborhood is determined by $p\textsubscript{stay}$ (Eq~\ref{eq:probstay}):

\begin{eqnarray}
\label{eq:probstay}
	\mathrm{p_{stay}} = \frac{\exp (\beta_{stay}(U-1))}{1+{\exp (\beta_{stay}(U-1))}}
\end{eqnarray}

\noindent
In this equation, $\beta\textsubscript{stay}$ is simply a calibrating factor that determines $p\textsubscript{stay}$. The choice to stay becomes deterministic in the limit $\beta\textsubscript{stay}\rightarrow\infty$, and completely independent of $U$ in the limit $\beta\textsubscript{stay}\rightarrow0$. For the simulations of the model, in order to represent real-world conditions, we choose a high value: $\beta\textsubscript{stay}$ = 1000.

Depending on the realization of $p\textsubscript{stay}$, if the agent action is an attempt to move from its current neighborhood, then this move occurs with probability $p\textsubscript{move}$, which depends on the difference between the wealth of the agent ($W\textsubscript{a}$) and the median wealth of the randomly selected receiving site ($W\textsubscript{R}\textsuperscript{Med}$). If $W\textsubscript{a}$ $\ge$ $W\textsubscript{R}\textsuperscript{Med}$, then the agent moves with probability, $p\textsubscript{move}$ = 1. (Previously~\cite{bib3aa}, we described in detail this inversion of the classic Schelling model~\cite{bib1} by replacing the tolerance level ($\tau$) of an agent with a threshold level (T) presented to all agents by extant neighborhood wealth configurations.) If not, the move occurs stochastically with probability $p\textsubscript{move}$ (Eq~\ref{eq:probmove}):

\begin{eqnarray}
\label{eq:probmove}
	\mathrm{p_{move}} = \exp (\beta_{move}(W_{a}-W_{R}^{Med}))
\end{eqnarray}

\noindent
Here, $\beta\textsubscript{move}$ is the calibrating factor that determines the probability of movement. In the limit $\beta\textsubscript{move}\rightarrow\infty$, $p\textsubscript{move}$ = 0 and the only moves that occur are those that satisfy the threshold condition: $W\textsubscript{a}$ $\ge$ $W\textsubscript{R}\textsuperscript{Med}$. Since a high $\beta\textsubscript{move}$ is potentially a reasonable representation of most real world cities, for the simulation of neighborhood economic status we keep $\beta\textsubscript{move}$ at the following values: 10, 100, 1000, and 100,000. Additionally, as we explored earlier~\cite{bib3aa}, decreasing $\beta\textsubscript{move}$ yields a progressive increase the number of moves where the threshold condition is contravened (disallowed-realized moves) and this in turn results in a corresponding non-linear increase in the number of moves satisfying the threshold condition (allowed moves). The overall effect of these dynamics is that there is a sharp transformation from a highly segregated to a mixed wealth configuration even at a very small fraction of disallowed-realized moves. Therefore, in addition to testing the dynamics of neighborhood economic status at high $\beta\textsubscript{move}$, we also seek to confirm that the sharp transformation from segregated to mixed wealth states still occurs with progressively decreasing $\beta\textsubscript{move}$. For this, $\beta\textsubscript{move}$ takes the following values: 10, 5, 2, 1, 0.1, 0.01, and 0.001

In order to model the relative economic status of neighborhoods and their changes over time, we use a simple stochastic wealth gain function for agents that is dependent on $U$. This probability of gain, $p\textsubscript{gain}$, is applied to each household that stays back in its current location, either because it chose to or because its attempt to move failed, and takes the same functional form as the function for $p\textsubscript{stay}$, and is given by (Eq~\ref{eq:probgain}):

\begin{eqnarray}
\label{eq:probgain}
	\mathrm{p_{gain}} = \frac{\exp (\beta_{gain}(U-1))}{1+{\exp (\beta_{gain}(U-1))}}
\end{eqnarray}

\noindent
Here, $\beta\textsubscript{gain}$ is a calibrating factor for $p\textsubscript{gain}$ and for the simulations, we use $\beta\textsubscript{gain}$ = 1. If $p\textsubscript{gain}$ is realized, then the household's wealth increases by a rate of $w\textsubscript{inc}$. For instance, if the wealth of a randomly sampled household is $W\textsubscript{a}$, and if $p\textsubscript{gain}$ is realized for this household, then the updated wealth of the household is given by (Eq~\ref{eq:wgain}):

\begin{eqnarray}
\label{eq:wgain}
	\mathrm{W_{a}} = (1 + w_{inc}) W_{a}
\end{eqnarray}

To summarize the update algorithm, at each iteration, a household is randomly chosen and it either stays back in its current neighborhood or attempts a move to a new neighborhood based on the Potential (U) of its neighborhood, which is manifested in the realization of $p\textsubscript{stay}$. If the realization results in an agent attempting a move, then this move occurs with probability $p\textsubscript{move}$. If the agent stays back in its current location (either because the realization of $p\textsubscript{stay}$ results in this option or because it attempted a move that failed), then it gains wealth with probability $p\textsubscript{gain}$. If $p\textsubscript{gain}$ is realized, then the agent wealth increases by a rate of $w\textsubscript{inc}$, which for the simulations is set to 5\%. If $p\textsubscript{gain}$ is not realized, then the agent wealth does not change.

The update process in the model is sequential (continuous), which means that all model variables are updated at the end of each iteration. The dynamics are run over 2,250,000 iterations for 45,000 agents (households), which means that on average, each agent is sampled 50 times. We define 45,000 iterations as 1 time sweep, which is essentially the number of iterations, on average, for every agent to have been sampled once. The 2,250,000 iterations therefore equate to 50 time sweeps of the dynamics. The choice of 50 time sweeps is based on the transient dynamics generated by the model, which is to say that it is at about 50 time sweeps that we observe outcomes that potentially correspond to Rosenthal's~\cite{bib11} empirical observations. We also then run the model further to 100 time sweeps to simulate even longer term outcomes and attempt to offer both predictions for longer-term changes in neighborhood Economic Status as well as possible explanations for the same. We run the entire model 25 times for each of the 8 combinations of $\beta\textsubscript{stay}$, $\beta\textsubscript{move}$, $\beta\textsubscript{gain}$, and wealth distribution - LN ($\mu$ = 0, $\sigma$ = 0.25) or LN ($\mu$ = 0, $\sigma$ = 0.5), enabling the creation of a reasonable ensemble of runs for each unique parameter set and a total ensemble of 200 model runs.

Given our objective, the parameters we seek to measure as outcomes over the 50 time sweeps are the relative economic status of neighborhoods and the extent of segregation.

Based on the work of Rosenthal~\cite{bib11}, we first define Economic Status ($ES$) of neighborhoods. The $ES$ of a neighborhood $C$ is the average household income of that neighborhood relative to the average household income of all neighborhoods in the city (Eq~\ref{eq:es}). 

\begin{eqnarray}
\label{eq:es}
	\mathrm{ES} \equiv \frac{W_{tot}(C)/P(C)}{\sum_{j=1}^{M} W_{tot}(j)/P(j)}
\end{eqnarray}

At the end of each time-sweep, the $ES$ of all the neighborhoods in the city is bunched into quartiles and each quartile is assigned into one of four Relative Economic Status ($RES$) categories - the first quartile (up to 25\textsuperscript{th} percentile) is RES I, the second quartile (between 25\textsuperscript{th} and 50\textsuperscript{th} percentile) is RES II, the third quartile (between 50\textsuperscript{th} and 75\textsuperscript{th} percentile) is RES III, and the fourth quartile (above 75\textsuperscript{th} percentile) is RES IV. 

Segregation ($S$) is measured as the fraction of population in neighborhoods (ranked in descending order of total neighborhood wealth) that own 80\% of the total wealth in the city. A low value of $S$ indicates greater segregation and a high value, a more mixed configuration.

Table~\ref{table1} summarizes the basic model parameters: \newline

\begin{table}[!ht]
\centering
\captionsetup{justification=centering}
\caption{\bf Model Parameters.}
\begin{tabular}{|l|l|}
\hline
Number of Neighborhoods ($M$) & 225\\ \hline
Number of Agents (Households) & 45,000\\ \hline
Number of Iterations & 2,250,000 \\ \hline
Number of Time Sweeps & 50 \\ \hline
$\beta\textsubscript{stay}$ & 1000 \\ \hline
$\beta\textsubscript{gain}$ & 1 \\ \hline
$\beta\textsubscript{move}$ & 0.001 - 100,000 \\ \hline
$w\textsubscript{inc}$ & 5\% \\ \hline
\end{tabular}
\label{table1}
\end{table}

This model derives its basic form from our previous model~\cite{bib3aa}, but it is however significantly extended by the addition of the concept of neighborhood Potential and the subsequent possibility of wealth gain. While the only decision an agent made in the previous model was whether to move, based on a simple of comparison of the median wealths of her current neighborhood and a randomly chosen alternate neighborhood, our current model describes a richer phenomenology incorporating the dynamics of agent choice, agent movement, and agent wealth gain. Also, the spatial structure of the models is different with the current model lacking any explicit spatial structure, while the previous model had individual cells representing individual agents, each with their own Moore neighborhoods. However, the findings of our previous model are robust even in the absence of explicit spatial structure, as demonstrated in S2 Appendix of~\cite{bib3aa}. We also use a log-normal distribution for the initial wealth distribution in this model as opposed to the normal distribution in the previous one. Again, the previous model is not contingent upon the usage of the normal distribution and as we mention in~\cite{bib3aa}, the model is robust to distribution choice. Therefore, even as we explore the dynamics of agent movement and wealth gain in yielding emergent neighborhood RES patterns in this model, we would continue to expect that the sharp transformation from segregated to de-segregated states observed in our previous model is replicated here as well.

\section*{Description of Emergent Dynamics}
\begin{figure}[!h]
\centering
\includegraphics[width=0.75\textwidth]{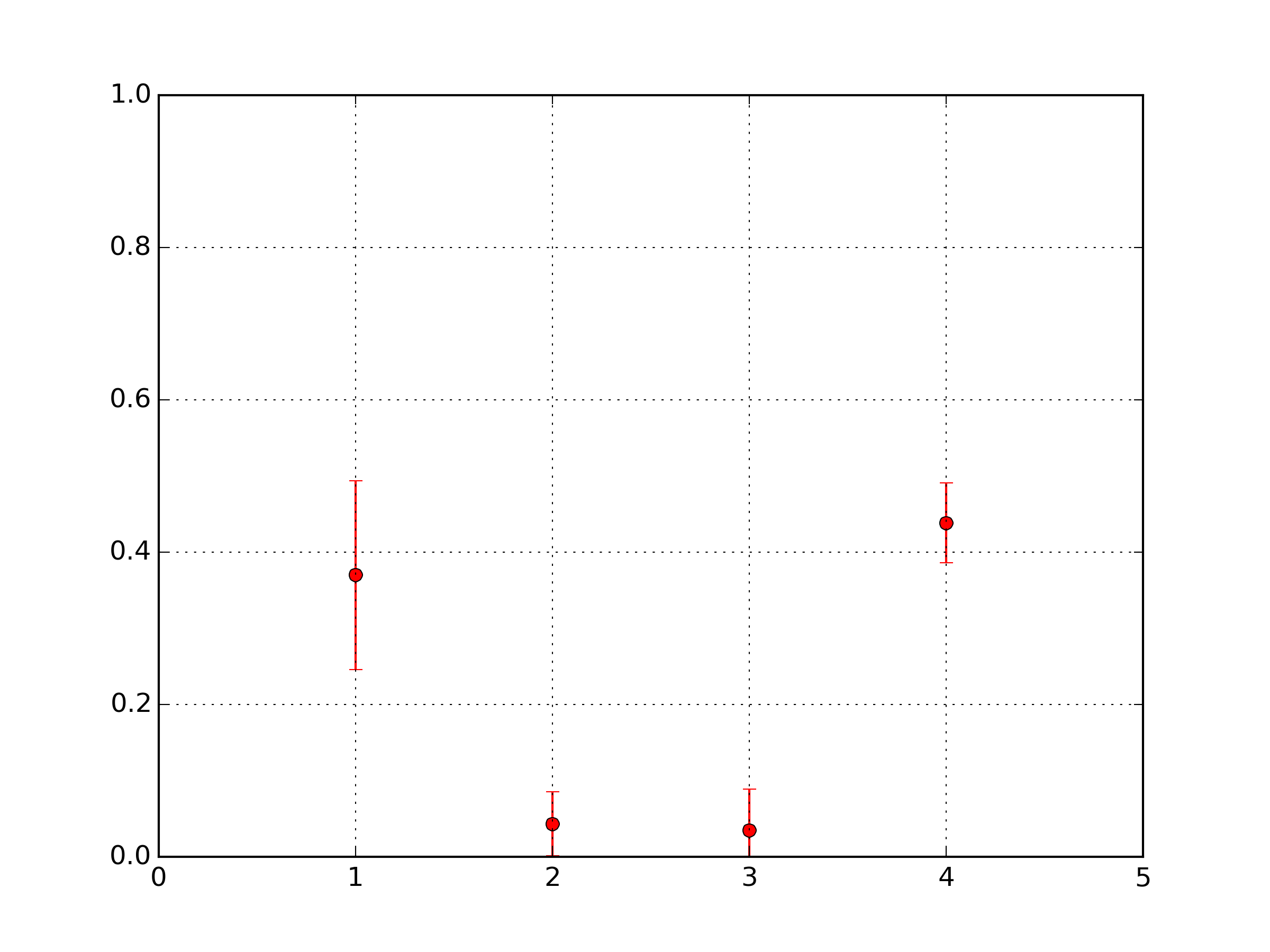}
\caption{{\bf Fraction of neighborhoods of different RES retaining status. }
Legend: Y-Axis: R = Fraction of neighborhoods retaining status; X-Axis: RES. 1 = RES I, 2 = RES II, 3 = RES III, 4 = RES IV.}
\label{fig1}
\end{figure}

The evolution of RES over time reveals that neighborhoods that start out in RES I and RES IV do indeed tend to more frequently retain status than neighborhoods starting out in RES II and III. Simulation results show us that, on average, RES I and RES IV neighborhoods retain status 37\% and 44\% of the time, while RES II and III retain status around 4\% of the time. The pattern of these outcomes is of a similar nature to the empirical outcomes observed by  Rosenthal in American cities~\cite{bib11}, where poor and rich neighborhoods retained status 34\% and 44\% of the time, while the lower-middle income and upper-middle income households retained status around 26\% of the time.  These patterns are displayed in Fig~\ref{fig1}, which is a plot of the averages of all 200 realized simulations of RES (with error bars). 

In addition to establishing the fraction of cells that retain RES status, we also track the amount of time (in number of continuous time sweeps) that a cell, on average, stays in a given RES so as to establish the time stickiness of each RES. Again, we find that when a cell enters RES I or RES IV it tends to stick for a larger number of time sweeps in these RESs, as against RES II and III, where it stays for fewer time sweeps before changing status. Specifically, we find that, on average, when a neighborhood enters RES I or RES IV, it tends to stay in these states for 14.3 and 17.7 contiguous time sweeps respectively; when it enters RES II or III, the average number of contiguous time sweeps spent in these states is 6.6 and 7.6 respectively. This further corroborates the evidence on the greater stickiness of states RES I and RES IV as compared to RES II and III.

\begin{figure}[!h]
\centering
\includegraphics[width=0.75\textwidth]{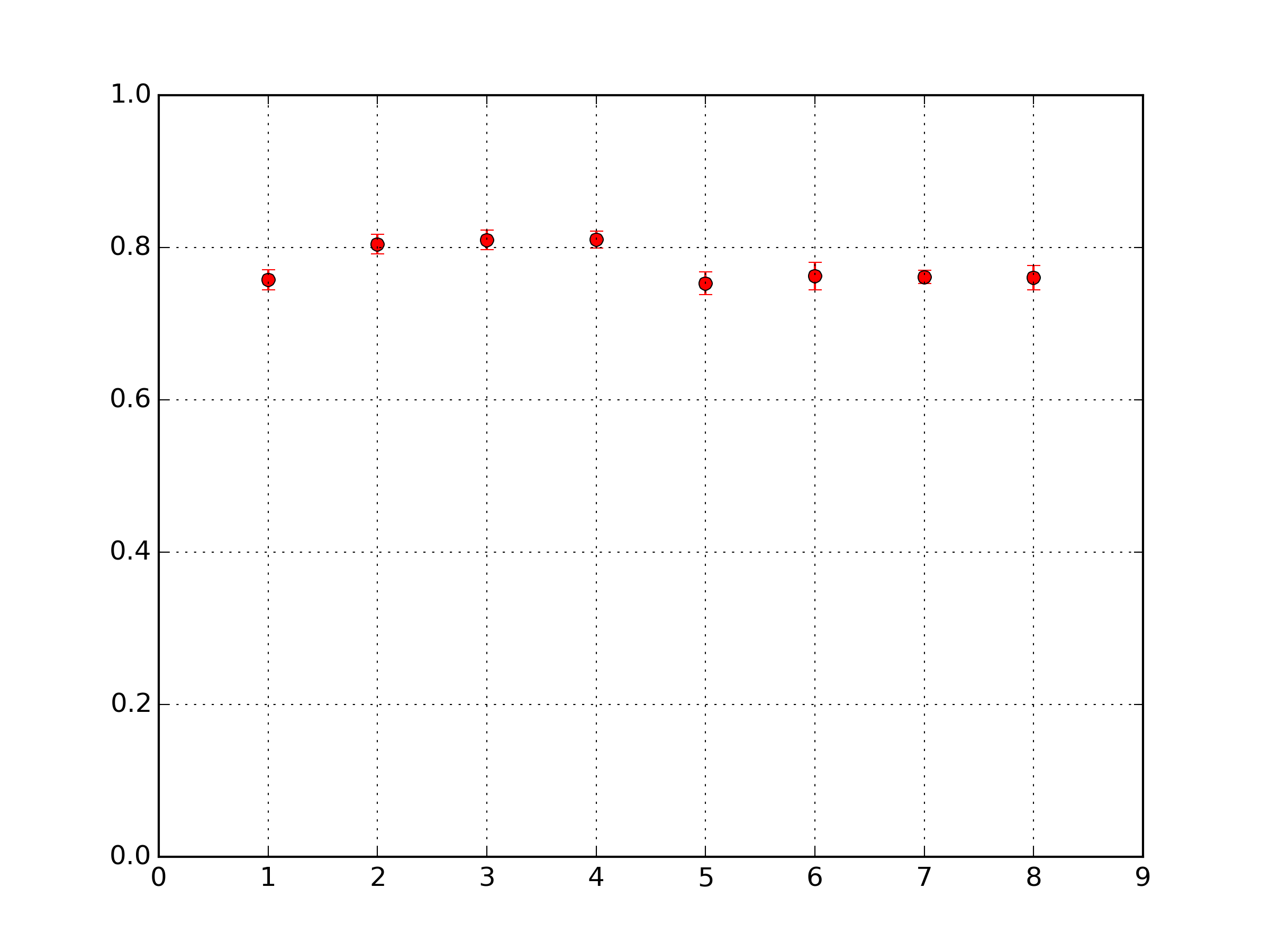}
\caption{{\bf Fraction of neighborhoods with changing RES for different runs. }
Legend: Y-Axis: F = Fraction of neighborhoods with changing RES; X-Axis: X = Run number. X = 1: $\beta\textsubscript{move}$ = 10, wealth configuration = LN ($\mu$ = 0, $\sigma$ = 0.25), F = 75.79\%; X = 2: $\beta\textsubscript{move}$ = 100, wealth configuration = LN ($\mu$ = 0, $\sigma$ = 0.25), F = 80.43\%; X = 3: $\beta\textsubscript{move}$ = 1000, wealth configuration = LN ($\mu$ = 0, $\sigma$ = 0.25), F = 81.00\%; X = 4: $\beta\textsubscript{move}$ = 100,000, wealth configuration = LN ($\mu$ = 0, $\sigma$ = 0.25), F = 81.05\%; X = 5: $\beta\textsubscript{move}$ = 10, wealth configuration = LN ($\mu$ = 0, $\sigma$ = 0.5), F = 75.31\%; X = 6: $\beta\textsubscript{move}$ = 100, wealth configuration = LN ($\mu$ = 0, $\sigma$ = 0.5), F = 76.27\%; X = 7: $\beta\textsubscript{move}$ = 1000, wealth configuration = LN ($\mu$ = 0, $\sigma$ = 0.5), F = 76.14\%; X = 8: $\beta\textsubscript{move}$ = 100,000, wealth configuration = LN ($\mu$ = 0, $\sigma$ = 0.5), F = 76.04\%.}
\label{fig2}
\end{figure}

The simulations also, on average, result in 78\% of the neighborhoods changing character in the course of 50 time sweeps. Rosenthal empirically observed \cite{bib11} that two-thirds of neighborhoods changed character over 50 years in American cities. Fig~\ref{fig2} plots the average fraction of neighborhoods changing RES for each of the eight parameter sets (with error bars).

\begin{figure}[!h]
\centering
\includegraphics[width=0.75\textwidth]{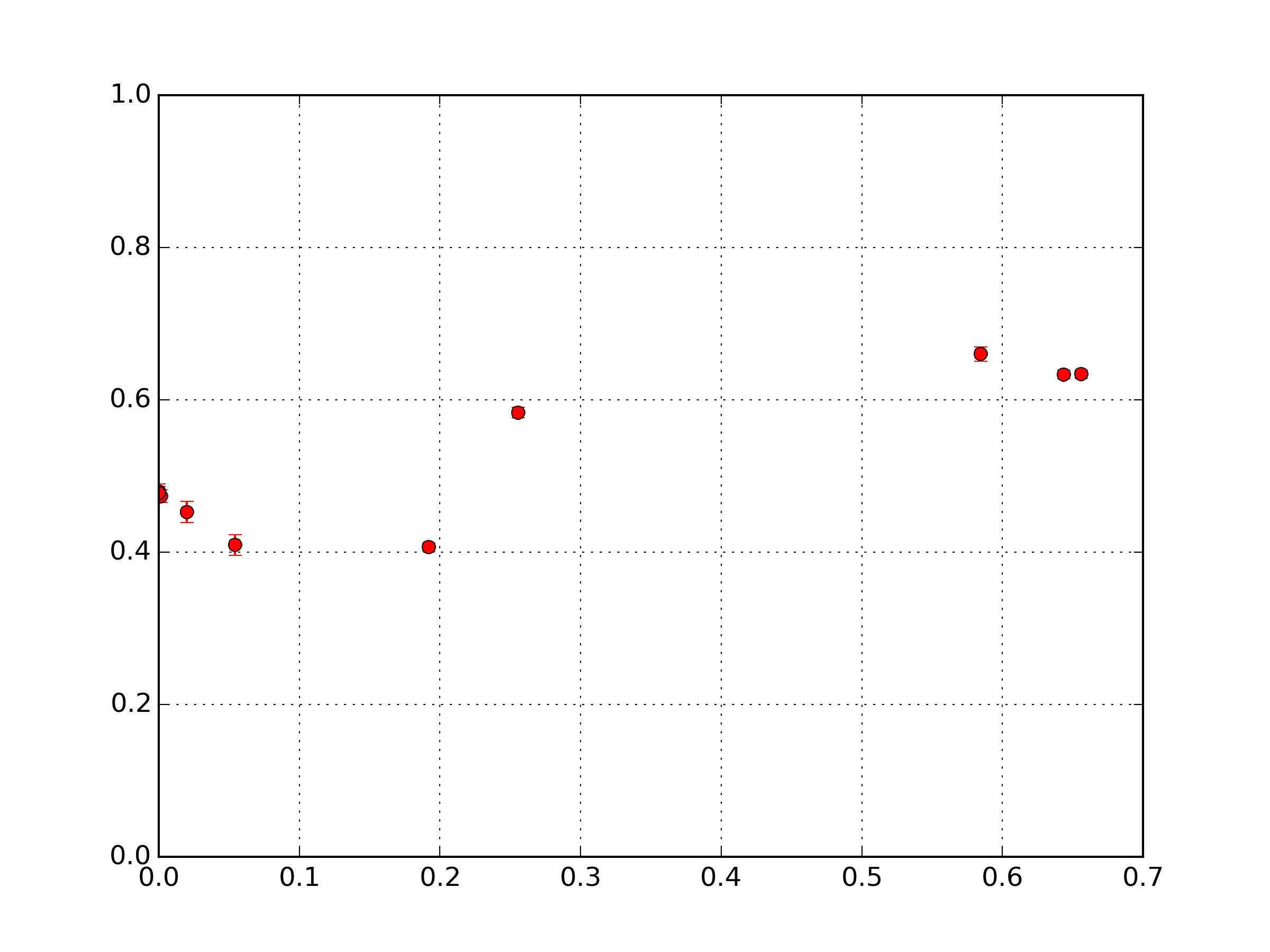}
\caption{{\bf Change in $S$ with ratio of disallowed-realized moves to attempted moves. }
Legend: Y-Axis: S = Fraction of population in the richest neighborhoods owning 80\% of total city wealth; X-Axis: Ratio of disallowed-realized moves to attempted moves.}
\label{fig3}
\end{figure}

Finally, we turn our attention to the onset of de-segregation with decreasing $\beta\textsubscript{move}$, and especially the sharp transformation between segregated and mixed states that was evident in the earlier version of this model~\cite{bib3aa}. We run the simulations for 100 time sweeps for wealth configuration LN ($\mu$ = 0, $\sigma$ = 0.25). We find, on average, over the 100 simulations that for $\beta\textsubscript{move}\geq2$, the measure of segregation, $S$, is in the region of 40-45\%, meaning that 40-45\% of the population in the richest neighborhoods owns 80\% of the city's wealth. However, as $\beta\textsubscript{move}$ drops below 2, we see a sharp increase in $S$, indicating a rapid decline in segregation. We interpret the impact of a declining $\beta\textsubscript{move}$ on $S$ in terms of the increase in disallowed-realized moves, and find that there is an essential drop in segregation (represented by a spike in the value of $S$) when the ratio of disallowed-realized moves to attempted moves is between 19\% and 25\%. It is in this range that $S$ increases sharply from 41\% to 58\% and ultimately appears to settle at a value close to 63\%. Fig~\ref{fig3} displays this sharp transformation from a state of high segregation to low segregation, confirming our earlier findings~\cite{bib3aa}.  

It is useful to highlight here that in the absence of wealth gain ($w\textsubscript{inc} = 0\%$), the dynamics do not yield segregation at all. This is on account of the inability of any cell to generate the momentum to pull its Potential away from the Potentials of other cells, even at high values of $\beta\textsubscript{move}$. This serves to therefore highlight the fact that the wealth gain mechanism, in addition to the Potential ($U$) function, is essential to the dynamics generating RES patterns.

\section*{Discussion}
We now explore the nature of the dynamics that result in the emergent RES patterns described earlier - for $\beta\textsubscript{move}\geq10$. Each agent's decision to attempt to move out of its extant neighborhood is based on the neighborhood's Potential ($U$), which is essentially a measure of its relative wealth. 

\begin{figure}[!h]
\centering
\includegraphics[width=0.75\textwidth]{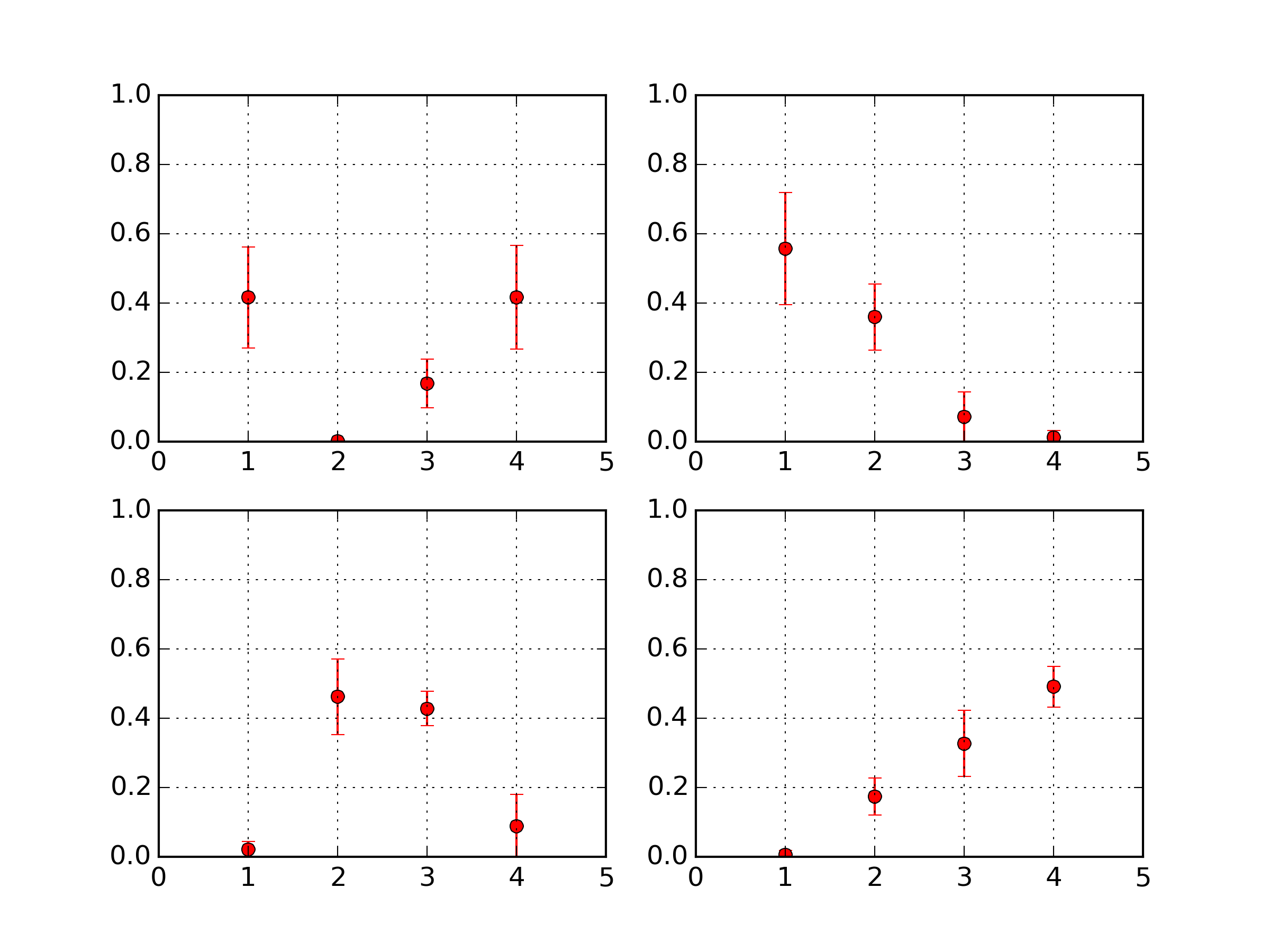}
\caption{{\bf Change in RES of neighborhoods by starting RES over 50 time sweeps. }
Legend: Y-Axis: Fraction of neighborhoods in given RES; X-Axis: RES. 1 = RES I, 2 = RES II, 3 = RES III, 4 = RES IV. Top Left: Starting RES = RES I; Top Right: Starting RES = RES II; Bottom Left: Starting RES = RES III; Bottom Right: Starting RES = RES IV.}
\label{fig4}
\end{figure}

For neighborhoods starting out with very high $U$ (RES IV), agents would prefer with very high probability to continue staying and gain wealth in these cells. This would mean that there is a strong tendency for RES IV neighborhoods to retain their character. However, there will be some small proportion of agents that indeed would stochastically choose to attempt and move out from these neighborhoods, resulting in a reduction in total wealth of these neighborhoods, and over time, potentially in their neighborhood RES. These opposing dynamics are clearly illustrated in Fig~\ref{fig4}, where we see that of all neighborhoods that start out in RES IV, on average (over all 200 ensemble runs), while 49\% end up in RES IV, a substantial 33\% and 17\% end up in RES III and RES II over the course of 50 time sweeps. Of the 49\% whose initial and final status is RES IV, close to 90\% have actually not changed their status at all in this time, indicating that given a cell is in RES IV, it remains in this state for long periods of time - fluctuation is unlikely. This confirms our intuition that RES IV neighborhoods tend to display a high level of stickiness and a significant fraction retain character; however the dynamics of a few agents in RES IV cells moving out of these neighborhoods does result in an equally significant proportion of these neighborhoods changing status to RES III and RES II over time.

Neighborhoods starting out with intermediate RES (RES II and RES III) tend to exhibit far greater change in RES. This is because, on the one hand, cells at RES II and RES III have values of $U$ that potentially result in a significant proportion of agents desiring movement to other locations (especially in RES II sites and to a lesser extent in RES III sites), which is matched on the other hand by a significant number of receiving cells with wealth thresholds conducive to accepting these agents. Fig~\ref{fig4} reflects this behavior as we see that, on average, 56\% of RES II sites end up as RES I sites and 46\% of RES III sites change into RES II sites. Additionally, Fig~\ref{fig4} significantly under-estimates the fraction of cells with fluctuating RES as it appears to indicate that 36\% of cells that start in RES II and 43\% that start in RES III also end in these respective states. We find that only about 10\% of the neighborhoods starting in RES II and RES III actually retain status over the entire 50 time sweeps; the rest of them have indeed seen fluctuations in their RES but have ended up at their initial RES at the 50\textsuperscript{th} time sweep. 

\begin{figure}[!h]
\centering
\includegraphics[width=0.75\textwidth]{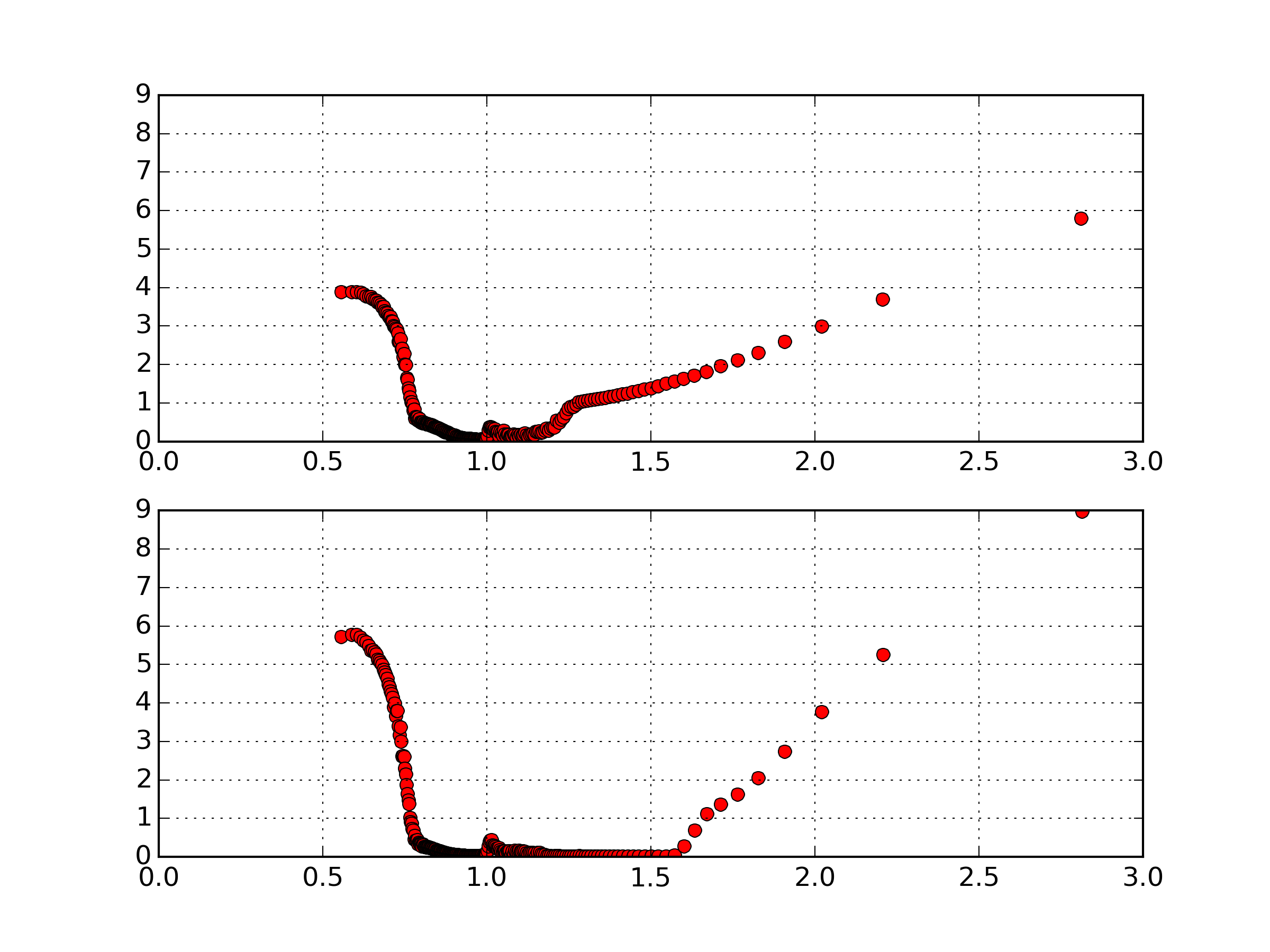}
\caption{{\bf Initial Economic Status v. Final Economic Status. }
Legend: Y-Axis: Final Economic Status of neighborhoods; X-Axis: Initial Economic Status of neighborhoods. Top: Status at 50 time sweeps; Bottom: Status at 100 time sweeps}
\label{fig5}
\end{figure}

In cells with low $U$, representing RES I neighborhoods, an agent would, with high probability, choose to attempt a move to a new location, but at $\beta\textsubscript{move}\geq10$, it will in most cases be unable to move to more wealthy neighborhoods (RES II, III or IV). Therefore, while this dynamic ensures that there would be a tendency for RES I neighborhoods to retain status, there are also countervailing dynamics that might be propelling RES I neighborhoods to higher wealth status over time. Namely, neighborhoods with very low U can also act as sinks for agents that move out of comparatively higher wealth neighborhoods. Since there may be agents in RES II/III/IV neighborhoods that choose to attempt moving out of their neighborhoods, the availability of low $U$ neighborhoods provides receiving neighborhoods with low threshold conditions to enable such moves. In the economics literature, such moves could be said to correspond to the process of gentrification~\cite{bib18} and over time due to the preferential attachment mechanism, could result in the gradual increase in the RES of these initially poor neighborhoods.  Fig~\ref{fig4} highlights both the increased tendency of RES I neighborhoods, on average 42\%, to retain status (and similar to the case of RES IV neighborhoods, close to 90\% of these neighborhoods have not changed status at all in this time) and also the consequence of preferential attachment dynamics that result in a significant proportion of RES I neighborhoods ending up as RES III (17\%) and RES IV (42\%) neighborhoods. This phenomenon of gentrification is quite clearly illustrated in Fig~\ref{fig5}, which represents, on average, the starting and ending RES of neighborhoods over 50 and 100 time sweeps (for the 100 ensemble runs of wealth configuration LN ($\mu$ = 0, $\sigma = 0.25$)). We see that there is indeed a fraction of cells starting at very low levels of Economic Status, which as a result of some initial fortuitous moves that generate the momentum for preferential attachment, end up with significantly increased Economic Status over 50 time sweeps. 

Given these dynamics, if we were to observe the neighborhoods over even longer periods of time than 50 time sweeps, we would expect that the behavior outlined above is further intensified for RES II, RES III, and RES IV neighborhoods - therefore resulting in very low levels of stickiness for RES II and RES III (close to 0\%), as well as a significant drop in RES IV stickiness. However, we would expect that the stickiness of RES I cells remains largely unchanged because the dynamics of preferential attachment have already been set in motion by the 50\textsuperscript{th} time sweep and it is unlikely that significantly more RES I cells are going to be able to graduate to higher RES. RES I neighborhoods that have retained status till the 50\textsuperscript{th} time sweep would therefore, with high probability, retain status over even longer time periods. This is confirmed when we run the simulations over 100 time sweeps, and find that the stickiness of RES I neighborhoods still remains, on average, 36.8\% (compared to 37\% at 50 time sweeps) but the stickiness of the other RES neighborhoods declines significantly. Stickiness of RES II, RES III, and RES IV neighborhoods now declines to, on average, 1\%, 1\%, and 20\%, as compared to 4\%, 4\%, and 44\% at the 50\textsuperscript{th} time sweep. Fig~\ref{fig5} provides further confirmation that RES I neighborhoods that benefit from the momentum generated by preferential attachment remain unchanged between 50 and 100 time sweeps. It is also apparent that the cells that had very low RES at 50 time sweeps still retain low RES at the end of 100 time sweeps.Therefore, the model essentially predicts that if the RES of neighborhoods in cities were observed for even more extended time periods, we would expect no stickiness for RES II and RES III neighborhoods, and low levels of stickiness for RES IV neighborhoods, while RES I neighborhoods display a much greater tendency to retain their levels of stickiness with time.

Additionally, the dynamics suggest that the nature of RES patterns is substantially different for increasingly mixed wealth configurations ($\beta\textsubscript{move}\leq2$) when compared to the more segregated configurations ($\beta\textsubscript{move}\geq10$). We find that for $\beta\textsubscript{move}\leq2$, over 50 time sweeps, there is, on average, no stickiness at all for RES I and RES II neighborhoods, negligible stickiness RES III neighborhoods ($\leq4\%$), and much higher stickiness in the region of 46\% for RES IV neighborhoods . This pattern suggests that, for $\beta\textsubscript{move}\leq2$, the dynamics enable sufficient disallowed-realized moves from RES I and RES II neighborhoods to create increasingly mixed wealth configurations, consequently ensuring that these lower RES neighborhoods are not stuck in long-term status traps. And as we have seen earlier in our review of literature on economic status of neighborhoods, long term economics status traps implying persistent segregation and continued economic decline can yield adverse impacts on education, employment prospects, civic participation, and mortality. What this highlights is the fact that if public policy measures could dynamically enable just enough disallowed-realized movement (mirroring $\beta\textsubscript{move}\leq2$), then we would see almost no stickiness of neighborhoods in low economic status (RES I) over time and this could potentially result in improved socioeconomic outcomes for larger fractions of households in cities. 

Overall, what this analysis highlights is the central role of the Potential ($U$) function in generating the emergent dynamics of neighborhood economic status. It is on the basis of $U$ that both agent movement and agent wealth gain are determined. When we run the dynamics by removing the Potential ($U$) and instead have agents randomly choose between staying to gain and attempting to move, we find that neither does segregation obtain even for  $\beta\textsubscript{move}\geq10$, nor do the neighborhood RES patterns match the empirically observed outcomes. We find that low RES neighborhoods tend to be extremely sticky  - with over 85\% of RES I neighborhoods retaining status and 50\% of RES II neighborhoods retaining status - while high RES neighborhoods exhibit no stickiness at all. The absence of segregation is explained by the fact that agents in rich neighborhoods are now equally likely to move as they are to stay, resulting in mixed configurations over long time periods. Cells that start in RES I will have half their agents, on average, staying to gain (with a random gain function) and the remaining trying to move out. The neighborhood threshold condition for movement ensures that only those with sufficient wealth are able to move to higher RES cells and for agents in RES I cells, most moves do not materialize initially. However, since wealth gain is also delinked from any notion of relative neighborhood wealth, the wealth gain of agents in a neighborhood becomes essentially random. As a consequence, over a period of time, many agents in RES I cells are able to gain wealth and move to other cells, but a small fraction of agents who have not gained, or gained enough to enable a move, remain back in these RES I cells. This explains the high levels of stickiness in RES I cells -  they have small, unwealthy populations. For cells starting out in RES III and RES IV, agents are able to gain randomly and move anywhere, and some of these cells see the congregation of large populations over time. The random process of wealth gain means that the status of most of these cells starting in RES III and RES IV keeps interchanging and as a consequence we observe no stickiness for these cells. This simulated outcome does not in any way replicate Rosenthal's~\cite{bib11} empirically observed outcomes.

The use of a metric such as Potential (U) in driving agent wealth gain and movement therefore appears critical in generating the dynamics which result in the emergence of segregation and apposite RES patterns. This also suggests that the patterns of RES empirically observed in American cities reflect an underlying universality and that we would indeed expect to see these repeated for cities around the world.

\section*{Conclusion}
We have here significantly extended our variant of the Schelling model to study the long term behavior of economic status of neighborhoods in cities. We add a Potential ($U$) function and a wealth increment function and find that the model reasonably replicates the empirically observed patterns of neighborhood economic status over long periods of time. Very rich and very poor neighborhoods tend to retain status more often than middle wealth neighborhoods. The use by agents of simple heuristics, such as the Potential ($U$) function, to compare the wealth of their neighborhoods relative to the wealth of the city in deciding if they want to move neighborhoods or stay back appears significant in driving the dynamics that yield the observed patterns of neighborhood economic status. We also find that our previous result on the sharp transformation from a segregated to a de-segregated state  still holds in the modified model.

\nolinenumbers

\end{document}